\documentclass[twocolumn,prd,showpacs,10pt]{revtex4}
\usepackage{graphicx}
\usepackage{dcolumn}
\usepackage{bm}
\def\lsim{\stackrel{<}{\sim}}
\def\gsim{\stackrel{>}{\sim}}
\def\wm{w_{m}}
\def\w0{w_{0}}
\def\at{a_t}
\def\dela{\Delta}
\def\s8{\sigma_8}
\def\LCDM{$\Lambda$CDM }
\def\be{\begin{equation}}
\def\ee{\end{equation}}
\def\bea{\begin{eqnarray}}
\def\eea{\end{eqnarray}}
\def\LL{\mathcal{L}}
\def\Om{\Omega}
\begin{document}
\title{The foundations of observing dark energy dynamics
with the Wilkinson Microwave Anisotropy Probe}
\author{P.S. Corasaniti${}^1$, M. Kunz${}^2$, D. Parkinson${}^3$,
E.J. Copeland${}^4$ and B.A. Bassett${}^{3,5}$}
\address{${}^1$ ISCAP, Columbia University, New York, NY 10027, USA}
\address{${}^2$ Astronomy Centre, University of Sussex, Brighton, BN1 9QJ, UK}
\address{${}^3$ Institute of Cosmology and Gravitation, University of Portsmouth, Portsmouth, PO1 2EG}
\address{${}^4$ Department of Physics and Astronomy, University of Sussex, Brighton, BN1 9QJ, UK}
\address{${}^5$ Department of Physics, Kyoto University, Kyoto, Japan}

\begin{abstract}
Detecting dark energy dynamics is the main quest of current dark energy research.
Addressing the issue demands a fully 
consistent analysis of CMB, large scale structure and SN-Ia data with
multi-parameter freedom valid for all redshifts. Here we undertake a ten
parameter analysis of general dark energy confronted with 
the first year WMAP, 2dF galaxy survey and
latest SN-Ia data. 
Despite the huge freedom in dark energy dynamics there are no new
degeneracies with standard cosmic parameters apart from a mild degeneracy
between reionisation and the redshift of acceleration, both of which
effectively suppress small scale power. Breaking this degeneracy will help
significantly in detecting dynamics, if it exists. Our best-fit model to
the data has significant late-time evolution at $z<1.5$. 
Phantom models are also considered and we find that the
best-fit crosses $w=-1$ which, if confirmed, would be a clear signal
for radically new physics.
Treatment of such rapidly varying models requires careful integration of the dark energy
density usually not implemented in standard codes, leading to crucial errors of up
to 5\%. Nevertheless cosmic variance means that standard $\Lambda$CDM models are
still a very good fit to the data and evidence for dynamics is currently
very weak. Independent tests of reionisation or the epoch of acceleration
(e.g. ISW-LSS correlations) or reduction of cosmic variance at large
scales (e.g. cluster polarisation at high redshift) may prove key in the
hunt for dynamics.
\end{abstract}

\maketitle

\section{Introduction}
Cosmological observations suggest that the Universe is dominated by
an exotic form of matter which is responsible for the present
phase of accelerated expansion \cite{PEL,ef,SPERGEL,Allen}. 
Several scenarios have been proposed to
account for the observations, but the nature of this
dark energy still remains unknown. The simplest minimal model to fit the experimental
data assumes the presence of a cosmological constant
$\Lambda$, representing the vacuum energy contribution to 
the spatial curvature of the space-time. In
spite of the success of this concordance $\Lambda$CDM
model, there is no reasonable explanation why the observed 
value of $\Lambda$ is extremely small compared to particle physics 
expectations \cite{Weinberg}. 

Alternatively, a light
scalar field, called quintessence,
rolling down a flat effective potential
has been proposed to account for the missing energy in the Universe
\cite{Wetterich,Ratra}. 
In particular quintessence models manifesting 'tracker' properties
allow the scalar field to dominate the present Universe 
independently of the initial conditions \cite{Zlatev,Steinhardt}.
Different realizations of the original quintessence idea have been
studied in the literature including the possibility of a scalar
field evolution driven by a non-canonical kinetic term \cite{Kess} and a non-minimal
coupling between quintessence and dark matter \cite{Coupl,Dom,Gaspe,Justin}. On the
other hand unified
models of dark matter and dark energy have been considered \cite{CHAPLYGIN,FREESE,CONDENSATION}. 
Despite the proliferation of scalar field models of dark energy,
we still lack of a fully consistent particle physics formulation.
Nonetheless distinguishing between a dynamical form of dark energy and a cosmological
constant is of immense importance as it would give
us a hint on the nature of this component.
The recent WMAP satellite measurements of the Cosmic Microwave Background, 
by providing an accurate determination of the anisotropy power
spectrum, offer the opportunity to have a better insight
into the physics of the dark energy. The quintessence hypothesis has been tested with
different methods using pre-WMAP CMB data and SN-Ia data or the 2dF power spectrum 
\cite{CORAS1,BRUCE,BACCI1,MELCH,MORT,LUCA}. 
These analysis have
constrained the dark energy equation of state $w$
without ruling out the possibility of a time dependence. 
In this article we carry out an analysis of the time evolution of the
dark energy equation of state. Our aim is to constrain
a set of parameters characterizing the dark energy properties and the standard cosmological
parameters by performing a likelihood analysis of the WMAP first year data \cite{BENNETT} and
the SN-Ia luminosity distance measurements \cite{PEL,RIESS04}.
The paper is organized as follows: in Section~\ref{met} and the
appendices we describe
the method and the data, in Section~\ref{like} we explain the evaluation
of the likelihood, in Section~\ref{res} we discuss our results
and finally in Section~\ref{conc} we present our conclusions.

\section{Method and data}\label{met}

\begin{figure}[ht]
\includegraphics[width=80mm]{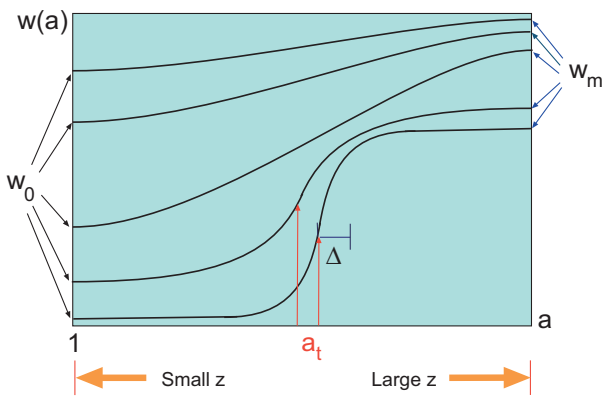} 
\caption[cpar]{\label{eosch} Schematic plot of the equation of state
paratrisation Eq.~(\ref{bcc}).}
\end{figure}

The most general way of constraining the time evolution of the dark energy equation
of state would require us to consider
a completely free function $w(z)$. As this corresponds to an infinite
number of new degrees of freedom, we have to simplify the problem.
We use a {\em physically} motivated parametrisation where $w = p/\rho$
is defined by its present value,
$\w0$, its value at high redshift, $\wm$, the value of the scale
factor where $w$ changes between these two values, $\at$ and the
width of the transition, $\dela$. Namely:
\begin{equation}
w(a)=w_0+(w_m-w_0) \Gamma(a, a_t, \Delta)\label{bcc}
\end{equation}
where $\Gamma$, the transition function, has the limits
$\Gamma(a=0)=1$ and $\Gamma(a=1)=0$ and varies smoothly between these
two limits in a way that depends on the two parameters $a_t$ and
$\Delta$ (see figure~\ref{eosch}).
Such a choice has been shown to allow adequate treatment of
generic quintessence and to avoid the biasing problems inherent in
assuming that $w$ is constant. Two choices for $\Gamma$ have been
given in the literature \cite{BRUCE,CORAS2} as discussed in Appendix
\ref{param}. Here we use the form advocated in \cite{CORAS2}. 

Using this general prescription has a profound advantage in attempts
to detect dark energy dynamics since, unlike simpler parametrisations
based on only one or two variables, it can accurately describe both
slowly and rapidly varying equation of states \cite{bck}. 
Detecting dark energy
dynamics and distinguishing it from a cosmological constant is
difficult and is clear only when there are rapid, late-time, changes
in $w$ \cite{CORAS3}, which then needs a formalism capable of
describing such rapid transitions. 

In order to compute the CMB power spectra, we use a modified 
version of the CMBfast Boltzmann solver \cite{ZALDA}. 

This is a non-trivial step in the case where $\Gamma$ changes rapidly (such as in 
our best-fit model!). In fact using a numerical method that is not able to track rapid transitions 
can lead to errors significantly larger than the error bars on the data, of order $5\%$, and consequently 
lead to completely wrong results. Our tests are described in detail in 
Appendix \ref{scaleq} and \ref{boltz}.

Several degeneracies amongst the cosmological parameters
prevent us from accurately constraining cosmological models
using CMB data only. Specific features of the anisotropy
power spectrum can provide information on particular combinations
of the cosmological parameters. For instance the relative height
of the Doppler peaks depends on the baryon density and the scalar
spectral index. In order to break such degeneracies it is necessary
to add external information. Since our goal is to 
constrain the properties of the dark energy, in a flat geometry
the main limitation comes from the geometric degeneracy between $w_0$, 
the dark energy density $\Omega_{DE}$ and the Hubble parameter $h$. 
This degeneracy can be broken by assuming an HST prior on the value of $h$
\cite{FREEDMAN} or/and combining the CMB with other data sets such as 
the matter power spectrum measurements from the 2dF galaxy survey 
\cite{Colless} or the SN-Ia data.
In our analysis we use the ``gold'' subset of the recent compilation of supernova
data of \cite{RIESS04} in addition to the WMAP TT and TE spectra.
An important point which we want to stress here is that CMB and SN data
can be treated at a fundamental level without any prior assumption
on the underlying cosmological model. For instance, this is the case for the matter power
spectrum data from galaxy surveys which implicitely assumes a \LCDM
model when passing from redshift space to real space. 
For this reason we add the 2dFGRS large
scale structure data only in order to check the 
stability of our results. We also remark that the use of
secondary observables such as the age of the Universe, the size
of the sound horizon at the decoupling, the clustering amplitude 
$\sigma_8$ (as quoted by the WMAP-team)
or the growth factor of matter density perturbations 
should not be used without thought to infer constraints 
on the dark energy since their quoted
value is usually derived by implicitly assuming a \LCDM cosmology.
This leads to biased results since these observables
depend on the nature of the dark energy \cite{Doran01a,Doran01b,CORAS3}. 
In principle CMB constraints can be easily added by using the position
of the Doppler peaks, which provide an estimate of the angular
diameter distance to the last scattering surface. But it is a well
known fact that pre-recombination effects can shift the peaks 
from their true geometrical position \cite{Doran01a}. 

The Integrated Sachs-Wolfe (ISW) effect also induces an additional shift 
in the position of the first peak, in a way that is strongly 
dependent on the evolution of the dark energy equation of state 
and is generally larger  than in $\Lambda$CDM models \cite{CORAS3}.
Therefore a consistent dark energy data analysis of the CMB indeed requires
the computation of the CMB power spectrum (and TE cross-correlation).

Each of our models is then characterised by the dark energy parameters
$\overline{W}_{DE}=(w_0,w_m,a_t,\Delta)$ and the cosmological parameters
$\overline{W}_{C}=(\Omega_{DE},\Omega_b h^2,h,n_S,\tau,A_s)$,
which are the dark energy density, the baryon density, 
the Hubble parameter, the scalar spectral index, 
the optical depth and the overall amplitude of the fluctuations
respectively. 
We therefore end up with ten parameters which can be 
varied independently. 

There is a remaining degeneracy in $n_S$, $\tau$ and $\Omega_b h^2$,
which allows the models to reach unphysically high values of the
baryon density and the reionisation optical depth. 
Following the WMAP analysis and in order to remain consistent
with \cite{Kunz}, we place
a prior on the reionisation optical depth, $\tau \leq 0.3$.
An alternative is to use a prior on $\Omega_b h^2$, which may
be physically better motivated and does a better job of breaking
the degeneracy -- we plan to use this for future work and discuss
the difference later on. We also limit ourselves to models with $w(z) \geq -1$, except where
stated explicitly.

\section{Evaluation of the likelihood}\label{like}

A grid-based
method would necessarily lead to a very coarse sampling, which is why
we opted for a Markov-Chain Monte Carlo (MCMC) method to sample
the likelihood surface. In addition
this approach also allows easy marginalisation over parameters.
We ran 16 to 32 independent chains on the UK national cosmology supercomputer
(COSMOS). This approach has both the advantage that there was no need
to parallelise the Boltzmann solver,
and that we are better able to assess the convergence and
exploration by comparing the different chains.

We take the total likelihood
to be the product of the likelihoods of each data set (CMB, SN-Ia
and LSS), or by defining $\chi^2_{\rm eff} \equiv -2 \log \LL$,
\be
\chi^2_{\rm tot} = \chi^2_{\rm WMAP} + \chi^2_{\rm SN1a} (+ \chi^2_{\rm 2dF}) ,
\ee
where the LSS contribution is added only in section \ref{lss}.
We evaluate the WMAP likelihood using the code of the WMAP science
team \cite{VERDE}, and we have checked that our SN-Ia likelihood
results are consistent with ones of Riess et al \cite{RIESS04}.
We treat the luminosity $\cal{M}$ as a nuisance parameter over
which we marginalise analytically. This automatically also
marginalises over the Hubble constant $H_0$, so that the supernova
data does not depend directly on it. For
the 2dF results we use the formalism and data of Tegmark
\cite{tegmark_2df}.

The convergence and sampling behaviour of the chains in a high-dimensional space
is far from trivial. Correlations in several dimensions are a particular issue as they lead to a
high rejection rate. To improve the acceptance rate, we estimate in a first
step the covariance matrix and then use rotated parameters which are linearly
independent (i.e. lead to a diagonal covariance matrix with variances 
$\sigma(p_i)^2$ on the diagonal). 
We set the proposal width of parameter $p_i$ ($i=1,..,N$) to 
$2.3 \sigma(p_i)/ \sqrt{N}$ as advocated in \cite{gilks,dunkley},
which we found to work very well. We also evaluate the rejection rate every 100
steps and adjust the width dynamically, but normally this is not necessary
once the covariance matrix is used.

The result of a MCMC is a ``cloud'' of samples in each chain with a density
proportional to the local value of the likelihood. In general we are
not content with a N-dimensional description of the likelihood, but would
prefer lower dimensional constraints. The generally accepted way to
project the likelihood onto fewer dimensions is through marginalising.
This means that we integrate the likelihood over the parameters which
we want to hide,
\bea
\LL(p_1,\ldots,p_{i-1},p_{i+1},\ldots,p_n)
&\propto& \nonumber\\
 \int dp_i \LL(p_1,\ldots,p_{i-1},p_i,p_{i+1},\ldots,p_n)&&
\eea

This requires a measure $\mu=dp$ on the parameter space. In many cases,
there is some physical motivation for the choice of the measure.
Alternatively, it may be that the for all 
reasonable choices, the measure does not vary strongly
across the range of interest.
As an example, we could use either $\Omega_b$ or $\Omega_b h^2$ as
a fundamental parameter. But since both $H_0$ and $\Omega_b h^2$
are quite well constrained, the result will not change appreciably
if we use the one or the other. In this case, it does not matter
which one we use. If neither is true then the result can depend strongly on the choice
of this measure. 

In the MCMC case, the choice of measure is implicit
in the choice of parameters if one follows the usual rule that the
marginalisation is done by summing up the samples. To illustrate
this, let us assume that the likelihood does not depend upon a
given parameter $0.1 \leq p \leq 10$. In this case, the marginalisation over this
parameter is just the volume of the parameter space, since all
values are equally likely. The resulting likelihood
will be flat, independent of the choice of parametrisation.
Using $p$ as our fundamental parameter, we find that roughly half the
points will be in $p<5$ and half in $p>5$. On the other hand,
using $\log p$ we will find that half the points are in $p<1$, and
half in $p>1$. The result depends in this case strongly on the
choice of parametrisation.
We would like to add here that a linear transformation of the
parameters does not change the measure, as long as the boundaries of
the integration are also adjusted. This is normally the case, since
the integration volume is usually given by the region where $\LL\neq0$.
An example of such a transformation is the use of a covariance matrix
to render the parameters linearly independent.

A different way to think about this specific problem, is by looking at
the choice of measure as a prior. Choosing for a parameter $p$ the
measure $dp$ (i.e. sample it equally in $p$) is the same as imposing
a flat prior, $P(p)=1$. Instead, sampling the parameter in $\log(p)$,
corresponding to a measure $dp/p$, corresponds to a flat prior in
$\log(p)$ or $P(p)=1/p$ when sampling evenly in $p$ \cite{tegmark_sdss}.

In order to test our dependence on the choice of parametrisation,
we ran chains for different choices, namely $(\at,\dela)$,
$(\log(\at),\dela)$ and $(\log(\at),\log(\dela))$. 
In the next section we will discuss which of the results depend on this choice.

A second issue which is often neglected is that the likelihood 
which we deal with here is by no means close to Gaussian in some of the variables.
The structure of the minima can thus be arbitrarily complicated.
It is therefore
important to specifically search for the global minimum, which can
of course be rather difficult in many dimensions.

As a final remark before discussing the results, we would like to point
out that caution is necessary when interpreting the results.
One sigma limits are not sufficient to rule out anything -- the true
model is indeed {\em expected} to lie about one sigma away from the
expectation value! Even two sigma or 95\% (the limits which we usually
quote) are not sufficient. We present here limits and constraints on
well over 20 different variables, and so we again expect at least one
to lie in the excluded region. Even worse, Gaussian statistics are well
known to underestimate the tails of ``real world'' distributions, so that
outliers are far more common than naively expected. In many fields, e.g. 
particle physics, a 5 sigma limit is being used to claim an actual
detection. Although our data is not yet of sufficient quality to impose
such stringent limits, we should bear this in mind. So we should as
an example only consider regions ruled out in the 1-dimensional likelihood
plots if the likelihood has (visually) fallen to zero.

\section{Results}\label{res}

In this section we discuss different aspects of our results. We start 
by taking a look at the best-fit models and at the goodness-of-fit
of both quintessence and \LCDM models. Then we show that the
introduction of a time-varying equation of state for the dark energy
component does not significantly alter the constraints on the basic cosmological
parameters. This allows us to discuss constraints on the time evolution of the
quintessence equation of state. In section \ref{lss} we use large-scale structure data instead
of the supernova data to break the geometric degeneracy. We also check
if the combination of both data
sets improves the constraints. Finally, we discuss limits on toy
``phantom energy'' models where we allow $w<-1$.

\subsection{The goodness of fit}\label{good}

Our global best fit QCDM model is characterized by the following dark energy parameters:
$\w0=-1.00$, $\wm=-0.13$,
$\at=0.48$ and $\dela=0.06$, which correspond to a fast 
transition in the equation of state at redshift of $1$. 
The total $\chi^2$ of this model is $1602.9$, while the 
best-fit \LCDM model has $\chi^2=1605.8$.
However, the total number of degrees of freedom is 1514, so that
all our fits are rather bad. This is mainly due to the
WMAP data (see the discussion in \cite{SPERGEL}).
In table~\ref{chi} we report the corresponding $\chi^2$ values for the CMB and SN data and
best fit values of the standard cosmological parameters for these two models. 
Notice that the QCDM model provides
the best fit to both the CMB and SN data.
\begin{table}[t]
\begin{tabular}{cccccccccc}
\hline
\hline
Model& $\chi^2_{CMB}$ & $\chi^2_{SN}$   & $\chi^2_{tot}$ & $\w0$ & $\Omega_{DE}$ & $H_0$  &  $\Omega_b h^2$ & $n_s$ & $\tau$ \\
\hline \\
$\Lambda$CDM & $1428.7$  &  $177.1$ & $1605.8$ & $-1.0$  &  $0.69$ & $69$ & $0.023$ & $0.97$ & $0.11$ \\
QCDM      & $1426.1$  &  $176.8$ & $1602.9$   & $-1.0$  &  $0.71$ & $67$ & $0.026$ & $1.09$ & $0.29$ \\ \\
\hline
\hline
\end{tabular}
\caption{\label{chi} $\chi^2$ and best fit values of the cosmological 
parameters for $\Lambda$CDM and QCDM models.}
\end{table}

\begin{figure}[h]
\begin{center}
\includegraphics[width=80mm]{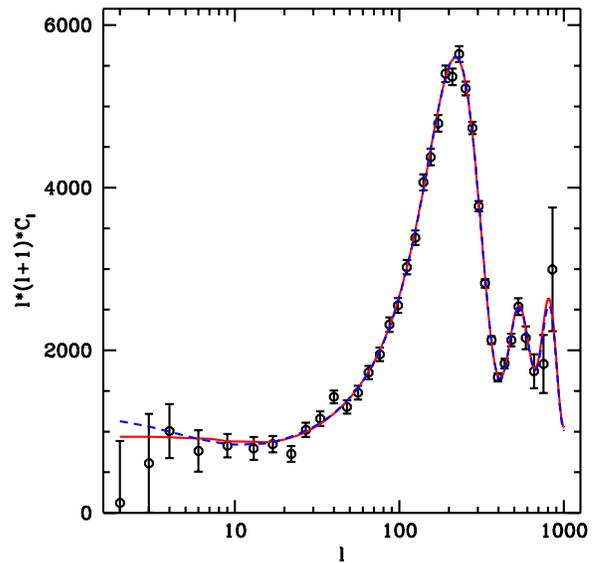} \\
\caption[cpar]{\label{cmb} CMB power spectrum for the QCDM (red solid line) and \LCDM (blue dashed
line) best fit models.}
\end{center}
\end{figure}

It is intriguing that such a model has a time evolving equation of state
$w(z)$ similar to that reconstructed from the best fit to the SN data in \cite{Alam,Peri,Cooray}.
In figure~\ref{cmb} we plot the temperature anisotropy power spectrum for these two
models. It is remarkable how perfectly the two completely different models agree
at intermediate $\ell$, demonstrating the power of the WMAP CMB data. At low multipoles
the additional freedom of the QCDM models allows a slightly better fit.
In fact due to a different distribution of the ISW,
these QCDM best fit models have less power at low multipoles than the \LCDM one.
However we want to remark that this part of the CMB spectrum is most likely
affected by galactic contamination effects \cite{Slosar}
and without a more accurate investigation 
we should not place too much emphasis on this suppression of power.
It is worth noticing that the difference in the best-fit cosmological parameters
between QCDM and \LCDM will lead to different TT power spectra at higher multipoles,
$l>700$. This suggest that an accurate detection
of the third peak may increase the statistical weight in favour or against the QCDM model.

The fact that the $\chi^2$ improves by $3$ through the addition of the
three new dark energy parameters, is to be
expected. Nevertheless we see that some quintessence models provide
a better fit to the data than standard \LCDM,
as opposed to analyses
which assume that $w$ is constant.
As we will discuss later, the limits on the time evolution of
the dark energy equation of state are still compatible with a number
of proposed scalar field scenarios. 

Studying the distribution of the $\chi^2$ values in the
MCMC chains for the \LCDM models, we find that 
$\Delta \chi^2 = 6.4$ for the models at $1\sigma$ (68.3\% CL)
and $\Delta \chi^2 = 11.8$ at $2\sigma$ (95.4\% CL).
In the Gaussian case, this corresponds to about
$5.5$ independent degrees of freedom, slightly less than
the number of cosmological parameters used ($6$).

For the quintessence models we find $\Delta \chi^2 = 9.9$ 
for $1\sigma$ and $\Delta \chi^2 = 15.5$ for $2\sigma$. 
Assuming Gaussian
errors, this would mean that we are dealing with
about $8$ independent degrees of freedom, many less
than the $6$ cosmological and $4$ dark energy parameters
used in the analysis.
We conclude that we are unable to constrain all the additional
parameters.

Is the improved $\chi^2$ of the 
best fit QCDM models a positive evidence for the additional
dark energy parameters? 
The answer to this question requires an estimation of
the information criteria associated with the model \cite{andrew_aic}. 
In general
adding more parameters tends to improve the fit to data, however
one should reward thoes models that can produce the same 
goodness-of-fit with fewer parameters. 
This can be achieved through the Akaike information criterion (AIC \cite{Akaike}) 
and Bayesian information criterion (BIC \cite{Bic}), respectively defined as
\begin{equation}
AIC=-2\ln{\mathcal{L}}+2k,
\end{equation}
and
\begin{equation}
BIC=-2\ln{\mathcal{L}}+k\ln{N},
\end{equation}
where $\mathcal{L}$ is the maximum likelihood, $k$ is the number of parameters of the
model and $N$ is the number of datapoints.
The \LCDM models have an AIC of $1617.8$, while the QCDM models of
$1622.9$. Based on this difference of $5$ we conclude that four parameter
dark energy models are not necessarily favoured. The BIC disfavours 
the new parameters even more strongly.
The information criteria therefore suggest that current data 
provide no positive evidence that the dark energy is
anything more complex than a cosmological constant.
This is not surprising, but should not be used as a reason
to stop searching for ways of detecting evidence of evolution. 
The rewards from finding such evidence would be huge.

\subsection{Constraints on cosmological parameters}

\begin{figure}[h]
\begin{center}
\includegraphics[width=80mm]{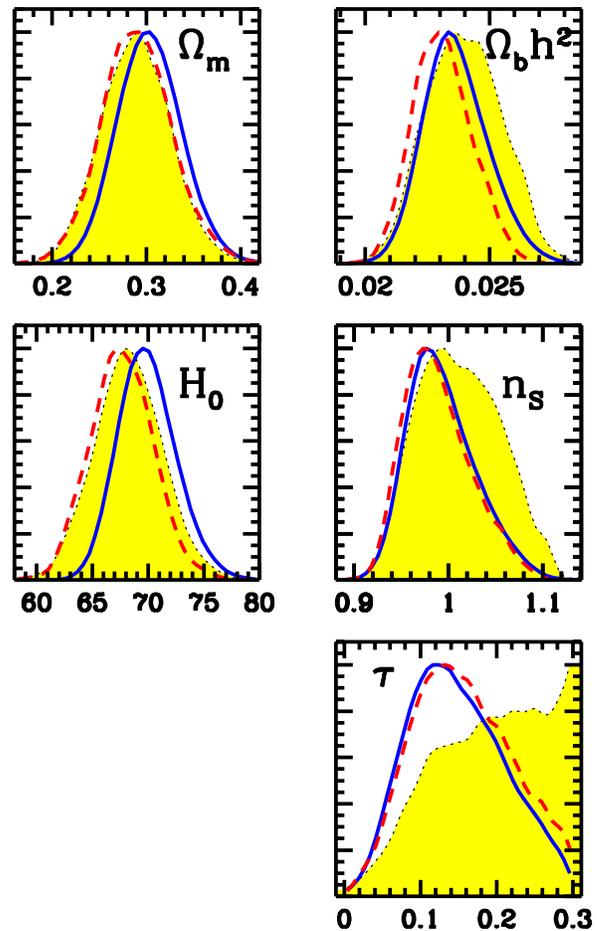} \\
\caption[cpar]{\label{cpar} Marginalized likelihoods for the various
cosmological parameters in the \LCDM scenario (blue solid curve) and including
the QCDM models (yellow shaded region). Also shown are QCDM models with 
a prior on the baryon energy density $\Omega_b h^2 = 0.0216 \pm 0.002$
(red dashed line). The results agree very well in all cases.}
\end{center}
\end{figure}

The class of quintessence models could {\em a priori} contain new, severe
degeneracies which change completely the preferred values of the
cosmological parameters. If this were the case, then all the standard
results of the WMAP analysis \cite{SPERGEL} would only be valid in
the context of a \LCDM model. We have found this not to be the case.
The new dark energy parameters
are degenerate amongst themselves but do not introduce any new degeneracy
with the other cosmological parameters. This can be seen in figure
\ref{cpar} where quintessence model results are compared with those obtained
for \LCDM models only. Unless specifically stated, the quoted results
have been obtained using a linear parametrisation of $\at$ and a logarithmic
one of $\dela$ in the MCMC. This maximises the weight of models with a late
rapid transition,
other parametrisations lead to an even slightly better agreement between \LCDM
and QCDM.

We notice that all parameters, with the
possible exception of the reionisation optical depth $\tau$, are
well determined in both cases. Also, their values are very similar
and cannot be distinguished even at one standard deviation.
Table \ref{tablep} lists the best fit values of the cosmological
parameters for the $\Lambda$CDM and the dark energy models.
The constraint on $h$ is consistent with the HST measurement \cite{FREEDMAN},
while the amount of clustering matter $\Omega_m$ is in agreement with
large scale structure estimates \cite{perci,allen}. The
physical baryon density is consistent with the Big-Bang Nucleosynthesis
expectations \cite{BBN}. This provides an important cross-check that
all viable models have to pass. The background universe is therefore
largely independent of the details of the dark energy.

\begin{table}[t]
\begin{tabular}{ccc}
\hline
\hline
Parameter      & $\Lambda$CDM        & QCDM  \\
\hline
\\
$\Omega_m$     & $0.30 \pm 0.03$     &  $0.29   \pm  0.04$   \\
$\Omega_b h^2$ & $0.0237 \pm 0.0013$ &  $0.0240 \pm 0.0015$  \\
$H_0$          & $70 \pm 3 $         &  $68     \pm 3$       \\
$n_s$          & $0.99 \pm 0.04$     &  $1.01   \pm 0.04$    \\
$\tau$         & $0.15 \pm 0.06$     &  $0.19   \pm 0.07$    \\
\\
\hline
\hline
\end{tabular}
\caption{\label{tablep}Mean and (formal) standard deviations for the cosmological
parameters. The $\Lambda$CDM values agree with published analyses.
The QCDM values are always consistent within one standard deviation,
showing that quintessence does not significantly impact ``precision
cosmology''.}
\end{table}

\begin{figure}[h]
\begin{center}
\includegraphics[width=80mm]{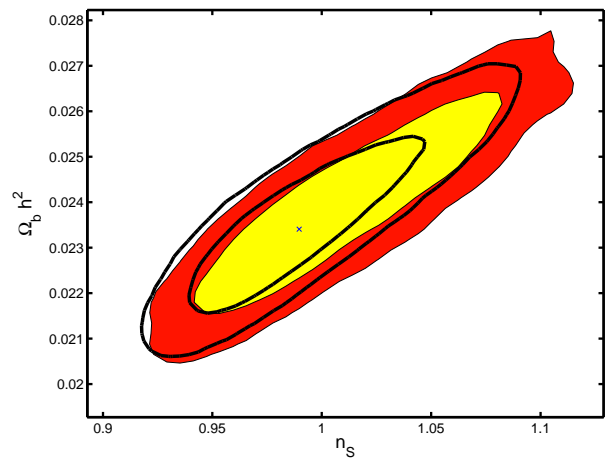} \\
\caption[cpar]{\label{ob_ns} The degeneracy between the scalar spectral
index, $n_S$, and the physical baryon density $\Om_b h^2$. The filled
contour are the 1 and 2 $\sigma$ limits of the quintessence models, while
the black solid contours are those of the \LCDM models.}
\end{center}
\end{figure}

\begin{figure}[h]
\begin{center}
\includegraphics[width=80mm]{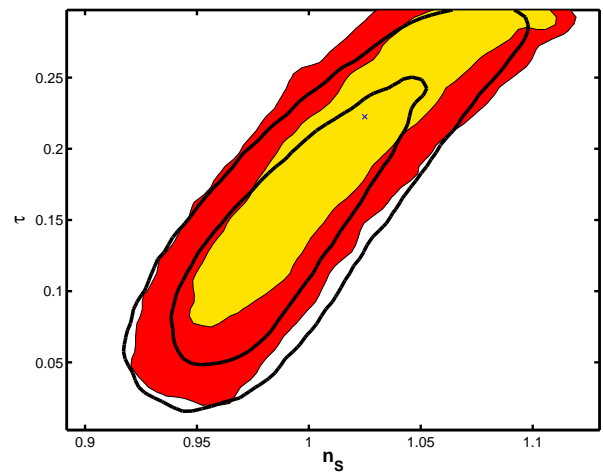} \\
\caption[cpar]{\label{ns_tau} The degeneracy between the scalar spectral
index, $n_S$, and the optical depth $\tau$. The filled
contour are the 1 and 2 $\sigma$ limits of the quintessence models.
The black lines show the corresponding limits for the \LCDM case.}
\end{center}
\end{figure}

We find nonetheless some differences, but they can be explained
quite easily. Firstly, these are the marginalised likelihoods.
The remaining degeneracy in $\Omega_m$, $H_0$ and $\w0$ is therefore
translated into a slight shift to lower values in both $\Omega_m$
and $H_0$. The degeneracy between the physical baryon
density, $\Omega_b h^2$, and the scalar spectral index, $n_S$,
becomes slightly worse (see figure~\ref{ob_ns}), leading to
somewhat longer tails in their likelihood distributions.
This is a consequence of the larger Integrated Sachs-Wolfe (ISW)
effect produced in time dependent dark energy models with the respect to the $\Lambda$CDM case \cite{CORAS3}. 
In fact the ISW boosts power on the large angular scales of the CMB,
therefore less power of the primordial
fluctuation power spectrum at small wavenumber $k$
is required to match the data. Consequently slightly
blue shifted spectral index values are preferred. The same effect
is also responsible for the spread of the likelihood distribution of the
reionisation optical depth through the degeneracy
between $n_s$ and $\tau$. For larger values of
$n_s$ the excess of power at the low multipoles of the TE spectrum requires
larger values of the optical depth (see figure~\ref{ns_tau}). Since the reionisation
suppresses the contribution at high $\ell$ with a factor of $e^{-2\tau}$ a better
determination of the CMB peak structure as well as better polarisation data will help
to break this degeneracy. We hope that the latter will be available
shortly, when WMAP releases the two-year data.

Figure \ref{cpar} shows also the
effect of introducing a prior on $\Omega_b h^2$, namely
$\Omega_b h^2 = 0.0216 \pm 0.002$. As we can see, the prior
only affects the cosmological parameters $\tau$, $n_S$ and
$\Omega_b h^2$ and removes some of their high values. None
of the dark energy parameters are affected (see figure~\ref{depar}). 
We should note that all cases use a top-hat prior $\tau < 0.3$.

\subsection{Constraints on the dark energy parameters}

As we have seen in the previous section, the additional parameters
which describe the dark energy do not introduce any new degeneracies with the standard
cosmological parameters. However we expect the dark energy parameter space to have an internal degeneracy.
For instance $\w0$ and $\wm$ both act on physical observables as an equation of state parameter.
Therefore slowly varying models with $\wm\approx\w0$ may look indistinguishable from
models with the same value of $\w0$ and a rapid transition at very early time from whatever value
of $\wm$.
As anticipated in Section~\ref{good}, the consequence 
of such degeneracy is that we can strongly constrain only one of the dark energy
parameters, which turns out to be $\w0$. 
In particular, the one-dimensional marginalized likelihood gives 
$\w0<-0.80$ at $2\sigma$, which is consistent with the 
upper limits quoted in other time dependent dark
energy analyses \cite{LUCA,DORAN,WANG}. Of course what this really implies 
is that the equation of state at a redshift of order z=0.1 is being constrained. 
We can not say anything about its true value today, 
but in what follows we use $\w0$ with this caveat in mind. 
The same applies when we take the limit of $z \to 0$ in the relevant figures.

The other dark energy parameters
are weakly constrained. In particular, as expected from the 
arguments discussed in Section~\ref{like}, 
we find that the inferred confidence intervals may depend 
on the parametrisation of $\at$ and $\dela$ used in the MCMC.
For the standard case, we find $w_m<-0.08$ at $2\sigma$, on the other hand
$\at$ and $\dela$ remain unconstrained. 
We refer to \cite{Kunz} for a discussion on breaking such a degeneracy
with an estimate of the value of $\sigma_8$ from large scale structure.

\begin{figure}[h]
\begin{center}
\includegraphics[width=80mm]{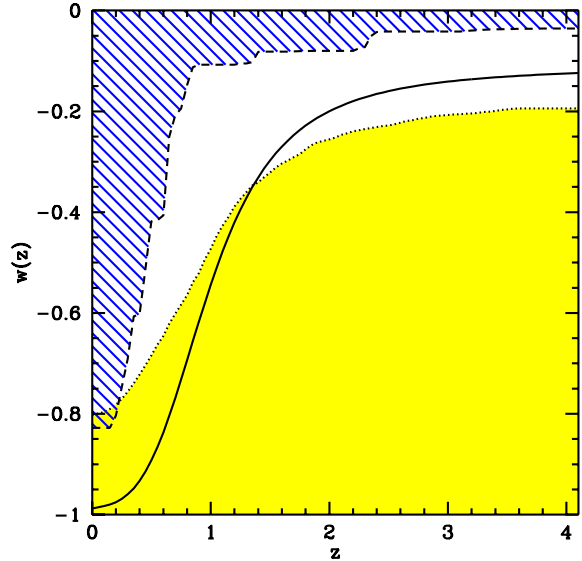} \\
\caption[cpar]{\label{wz} $2\sigma$ confidence region on 
$w(z)$ derived by taking the 95\%
models with lowest $w(z)$ from our main chain (yellow shaded area) and 
the ``exclusion zone'' where {\em all} models have $\Delta\chi^2 > 4$
from the best-fit model in our main chain (blue hatched area). We also
show the $w(z)$ of the best-fit model (black solid line). \LCDM is
acceptable at $2\sigma$.}
\end{center}
\end{figure}

A parametrisation in
$\log(\at)$ would place more emphasis on early transitions, 
so that the effective redshift where $w_m$ is evaluated is
moved to higher values. In this case the limits on $w_m$
become even weaker, and we conclude that $w_m$ as a parameter
is difficult to interpret. On the other hand, the limits on
$w(z)$ as a function of redshift are less dependent of the 
parametrisation of $\at$, as we derive them at all redshifts separately.
We therefore advocate these limits, as plotted in figure
\ref{wz}, as a better illustration of the constraints on
dark energy models.

\begin{figure}[h]
\begin{center}
\includegraphics[width=80mm]{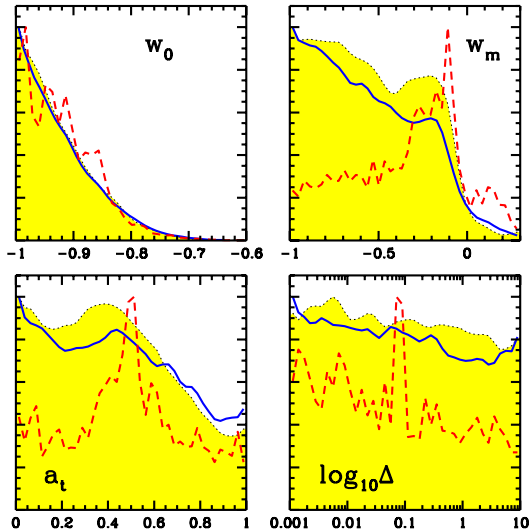} \\
\caption[cpar]{\label{depar} Likelihoods for the 
dark energy parameters in the QCDM models without prior on
$\Omega_b h^2$ (yellow shaded area) and with the prior (solid
blue line). The prior does not affect the dark energy parameters
significantly. We also show the relative goodness of fit of the
best model in each bin (red dashed lines).}
\end{center}
\end{figure}

These limits, derived with the Markov-Chain approach, rest solely
on the local density of the accepted models. If the assumption
of Gaussian errors holds approximately, then we can derive the
same limits using the actual likelihood values instead. In this
case, and considering a single variable, models with 
$\Delta \chi^2 < 4$ occur 95.4\% of the time. We find that
models with $\w0 >-0.8$ have $\Delta \chi^2 > 4$, consistent
with the limits from the Markov Chain.
In order to put limits on the equation of state as a function of redshift,
we proceed as follows: 
we compute $w(z)$ for each model
and then take either the $95\%$ confidence region over all models
in the chain (shaded area in figure~\ref{wz}) or compute the
highest $w(z)$ for models with $\chi^2 < \chi^2_{\rm min}+4$ (dashed line).

The shaded area is {\em a priori} the proper marginalised result,
representing the $95\%$ limits on $w(z)$.
However it is worth remarking that this constraint may suffer
from a potential problem with the choice of the
measure that we introduce to integrate the marginalised parameters over 
(see discussion in section \ref{like}). On the other hand  the
dashed line relies only on the goodness of fit, and can be
interpreted as an exclusion limit. In other words models with a $w(z)$ 
that enter the hatched area are a ``bad'' fit to the data. This
does not include any information about how likely these models
are, given the variation in the other parameters.
Our interpretation of this is as follows: if we want to judge if a
single, specific model is ruled out or not, then we should
be using the dashed line as an upper limit for $w(z)$. 
If we are more interested in
what we would expect as the value of $w(z)$, given our additional
knowledge about the other variables, then we should look at the shaded
region as providing a limit on the equation of state parameter.
As we can see in figure \ref{wz}, models with $\wm\geq 0$
at $z>1$ and with a fast transition occurring
at $z\lesssim1$ are
a bad fit to the data, being beyond the dashed blue line. 
Physically this is because models where the transition from $w_m$
to $w_0$ occurs at redshifts $z<10$ with $w_m>-0.1$ give rise to a non-negligible
dark energy energy contribution at  decoupling, which is strongly constrained
by CMB data. However, models with $\w0<-0.8$ and $\wm>-0.1$ for which
the transition occurs at redshift $z>10$ are consistent with the 
data as their early energy contribution is negligible. As mentioned
earlier, the apparent exclusion of these models based on the likelihood
for $w_m$ in figure~\ref{depar} is an artifact of our parametrisation of
$\at$. This can also be seen by noticing that the maximised results
(dashed curve) does not fall to zero for $\wm>0$, indicating that there
are acceptable models in this part of the parameter space.
However these models will be indistinguishable from a pure $\Lambda$CDM scenario
and are thus not very interesting when we try to rule out one from
the other.

\begin{figure}[h]
\begin{center}
\includegraphics[width=80mm]{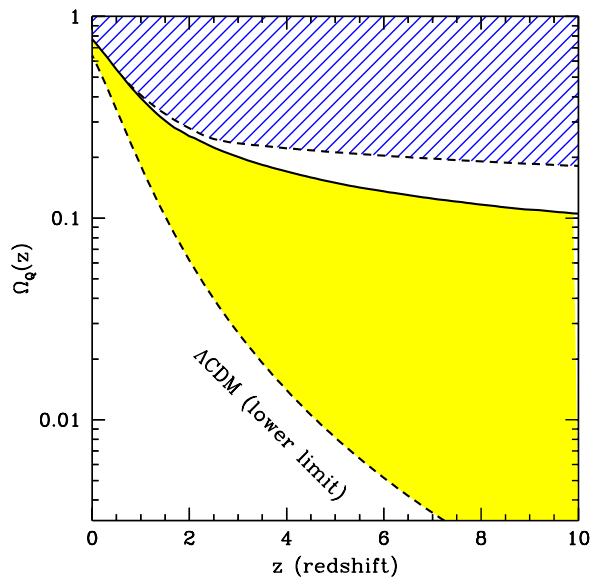} \\
\caption[cpar]{\label{Oqz} Limits on $\Omega_{DE}(z)$ corresponding
to those on $w(z)$  in figure \ref{wz}.}
\end{center}
\end{figure}

The limits on $w(z)$ can be reinterpreted as constraints on the
evolution of the dark energy density (see figure~\ref{Oqz}). Models
with $\Omega_{DE}(z)$ above the dashed line are ruled out, thus limiting
the amount of dark energy available during matter domination to be $\Omega_{DE}(z)<0.1$ (see also \cite{DORAN,Doran}).

These results undoubtedly have an effect for quintessence model building, and we will be investigating this aspect in a future paper.

For now though we restrict ourselves to a few general remarks. The limits on $w(z)$ previously obtained
allow us to constrain a large class of quintessence models.
For instance the exclusion plot in figure \ref{wz} suggests that models with a 
perfect tracking behavior, for which $w=0$ during the matter era up to $z \leq 4$, and
with a late time fast transition, are disfavoured by the data. As we have seen before,
this is because $\Omega_Q$ must be negligible at early times.
This class of models, for some particular choice of the parameters of the scalar field 
potential, include the two exponential potential \cite{NELSON} and
the Albrecht-Skordis model \cite{ALBRECHT}. Of course, 
if they leave their tracking behaviour before then, the constraint is weakened. 
These models satisfy the constraints only if the slope of the scalar field potential,
where the quintessence is initially rolling down, is very steep and then followed
by a nearly flat region such that the equation of state $\approx-1$ at the
present time. On the other hand models with a non-perfect tracking behavior 
and a slowly varying equation of state with $\w0<-0.8$ are consistent with the data.
This is the case of quintessence models with an inverse power law potential \cite{Ratra},
supergravity inspired potentials such as the one proposed in \cite{BRAX} or
off-tracking quintessence models, such as those studied in \cite{Axel,Kneller}.
Models of late time transition \cite{PL} which show features of
our best fit model can also be consistent with the data.

\subsection{Adding large scale structure\label{lss}}

\begin{figure}[h]
\begin{center}
\includegraphics[width=80mm]{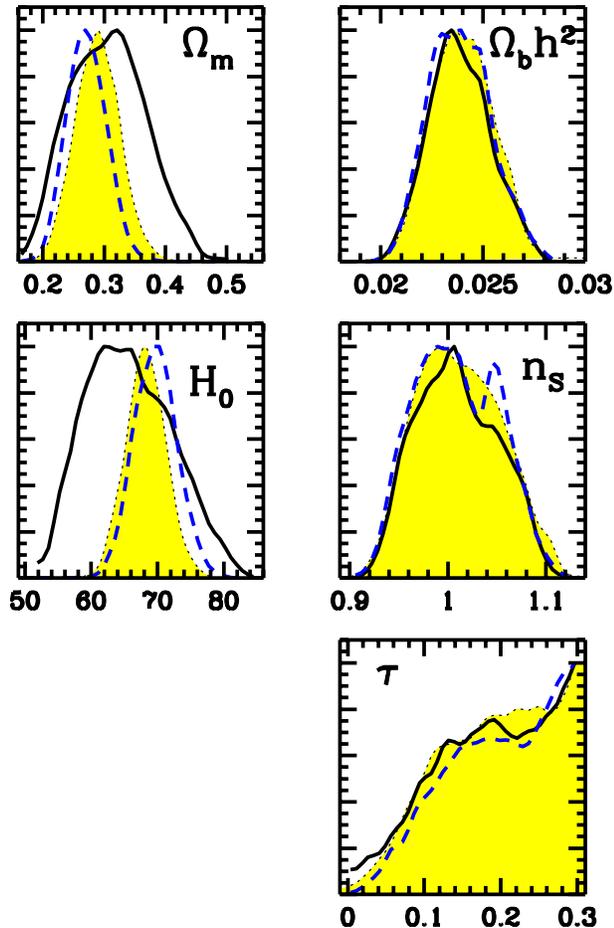} \\
\caption[cpar]{\label{cpar_lss} Constraints on the cosmological parameters
when adding the 2dFGRS data to CMB and SN-Ia data (blue
dashed line) and when only using CMB+2dFGRS (black solid line), compared
to the CMB+SN-Ia case (yellow shaded region). }
\end{center}
\end{figure}

\begin{figure}[h]
\begin{center}
\includegraphics[width=80mm]{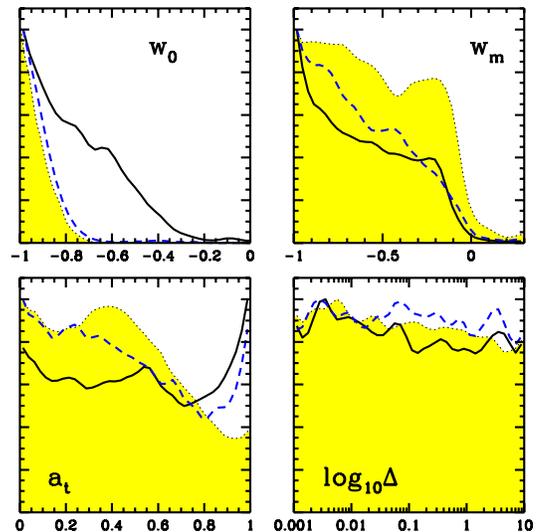} \\
\caption[cpar]{\label{depar_lss} Constraints on the dark energy parameters
when adding the 2dFGRS data to CMB and SN-Ia data (blue dashed
line) and when only using CMB+2dFGRS (black solid line), compared
to the CMB+SN-Ia case (yellow shaded region). }
\end{center}
\end{figure}

One important question is whether the supernova data contains
any severe systematic effects that may strongly bias our results.
To test for this possibility, we replace the supernova data by 
the 2dF galaxy redshift survey
power spectrum from ref.~\cite{tegmark_2df}. The bias is added as
a new, free parameter. This means that only the form of $P(k)$
is constrained, not its amplitude. The solid lines in figures
\ref{cpar_lss} and \ref{depar_lss} show the resulting likelihoods of the 
cosmological and the dark energy parameters respectively. Comparing
them to the CMB+SN-Ia results (shaded yellow regions), we find that
the results are consistent, but that the supernovae allow us to place
stronger constraints on $\Omega_m$, $H_0$ and $\w0$ when combined with
the CMB. There is no evidence for any systematic problems in the
SN-Ia data set.

Additionally, we can use the combination of all three data sets. 
In \cite{BRUCE} we found that the large scale structure (LSS) does not
add any strong constraints, beyond those found with CMB+SN-Ia data,
as long as no constraints on the bias parameter are imposed.
As the dashed curve in figures \ref{cpar_lss} and \ref{depar_lss}
shows, this is still the case, and the bias
parameter is strongly correlated with the clustering strength,
$\sigma_8$. 
The constraints on the bias found in \cite{POPE}
do not apply to our analysis, since they were obtained
by combining the SDSS 3D matter power spectrum with the WMAP 
results on $\sigma_8$ which are correct only for $\Lambda$CDM cosmologies.

For standard quintessence models one can thus either use supernovae
or large scale structure data, with the supernova data giving the
stronger constraints. The situation changes if additional
parameters need to be constrained, e.g. non-zero neutrino masses.
In this case it is crucial to have very good clustering data on small
scales where the neutrinos impose a distinct signal on $P(k)$.

\subsection{Phantom energy models}

Several authors have suggested that dark energy models with
a super-negative equation of state, for which $w<-1$, can
provide a better fit to the CMB and the SN-Ia data
\cite{CAL,MORT,TROD,KamCald,Jochen}. On the other hand
all these analyses are biased in favour of phantom 
dark energy models since they use
a constant equation of state parameter \cite{MAOR,VIREY}.
In this scenario the dark energy violates the weak energy condition (WEC)
which leads to a number of problems \cite{CAR}. For this reason
we feel that these models are disfavoured on theoretical grounds.
Nevertheless it remains an interesting question whether they
are compatible with the current cosmological observations?
Constraining time dependent phantom energy models allows us the opportunity
to test these models against the observational data without
the bias induced by assuming a constant equation of state parameter.
However the main problem for this class of models is the existence
of tachyon instabilities which lead to an exponential growth of phantom
energy perturbations on small scales. 
In addition
our formalism does not allow us to follow the evolution of the dark energy perturbations
for phantom models which cross the $w=-1$ value (see discussion in Appendix~\ref{scaleq}).
On the other hand we can account for the perturbations in models which violate the weak energy condition at all times.
 
\begin{figure}[h]
\begin{center}
\includegraphics[width=80mm]{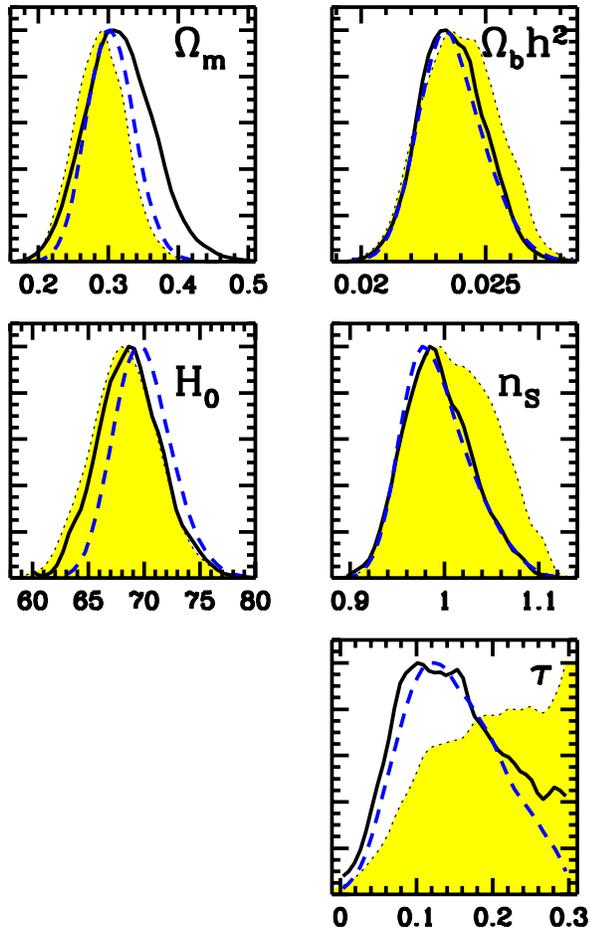} \\
\caption[cpar]{\label{cpar_phantom} Constraints on the cosmological parameters
for all phantom models (solid lines) and those that respect the
weak energy condition ($w\geq1$) (shaded), compared to the
\LCDM models (dashed).}
\end{center}
\end{figure}

In order to be self-consistent we therefore extend our analysis to two different
classes of dark energy models, those which always satisfy/violate the WEC and those
which cross the WEC boundary value $w=-1$. 
The latter are assumed to be 
homogeneous and consequently our analysis for these class of models accounts only
for the effects they produce on the background expansion.
Although this is not physical, we are unaware of a unique
prescription for handling these cases. We therefore suggest that the reader sees this section
more as a speculative treatment.
It is interesting to note that in these ``toy'' phantom models, 
the allowed values of the cosmological parameters do not change very much, in that they lie somewhere
between the \LCDM and the QCDM case (see Fig.~\ref{cpar_phantom}).

\begin{figure}[h]
\begin{center}
\includegraphics[width=80mm]{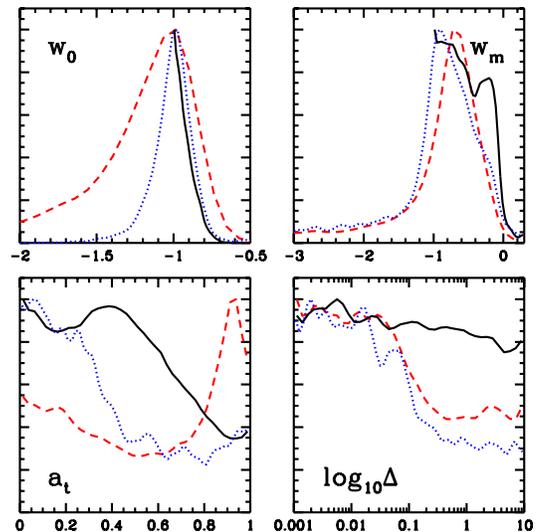} \\
\caption[cpar]{\label{depar_phantom} Constraints on the dark energy parameters
for all phantom models (red dashed lines) and those that respect the
weak energy condition ($w\geq-1$) (solid black line). The dotted line
is from models that do not cross $w=-1$.}
\end{center}
\end{figure}
The models which cross $w=-1$ provide a slightly improved best-fit.
In particular we find a best-fit model with 
$\chi^2=1601$. It has $\w0 = -2.0$ and
$\wm = -0.7$, i.e.~the equation of state crosses over that of the
cosmological constant, $p=-\rho$. This behaviour is most likely driven
by the supernova data, as we find a similar result if we only use
this data set \cite{bck}.

\begin{figure}[h]
\begin{center}
\includegraphics[width=80mm]{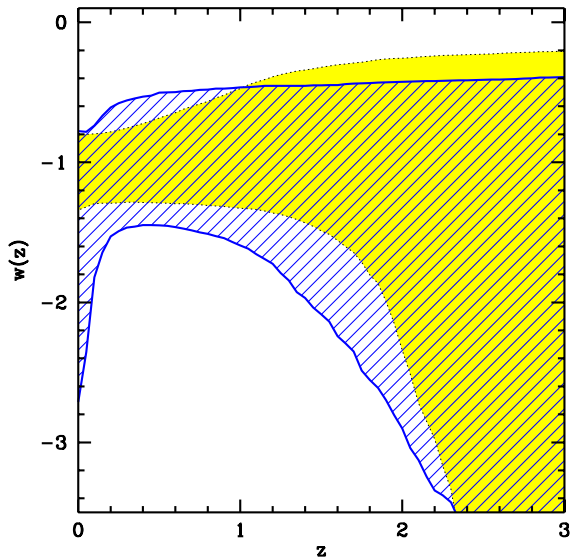} \\
\caption[cpar]{\label{wlim_phantom} Constraints on the dark energy equation
of state parameter $w(z)$. The shaded area corresponds to models which do not cross $w=-1$ and for which
perturbations are taken into account, while the hatched area corresponds to all models, including those
crossing $w=-1$, but without perturbations.}
\end{center}
\end{figure}

For the standard parametrisation of $\at$ and $\dela$,
the median for $\w0$ is $-1.1$ and for $\wm$ is $-0.8$. The
95\% confidence intervals are $-2.7 < \w0 < -0.77$ and $-8.3 < \wm < -0.19$.
For the models that do not cross $w=-1$, but 
always remain either with $w < -1$ or $w\geq-1$,
the results give the median for $\w0$
to be $-1.0$ and for $\wm$ to be $-0.97$. The overall best-fit model is the
same as for the standard QCDM models, but as the above median values
show there are about the same number of accepted models on both sides
of the divide.

Thus there is no evidence for any deviation from \LCDM in this extended framework.

The parameter which does show some change is the clustering
amplitude, $\s8$, which lies now in the 95\% confidence interval
$0.66 < \s8 < 1.24$ as opposed to the dark energy models satisfying the weak energy condition
$w\geq-1$ for which $0.53 < \s8 < 1.07$.

The limits on $w(z)$ are shown in figure~\ref{wlim_phantom}.
The shaded regions correspond to all models which do not cross $w=-1$ and for which
perturbations are included, while the hatched area corresponds to all models which cross $w=-1$.

\section{conclusions and outlook}\label{conc}

In this paper we have made the first attempt to 
constrain dynamical models of dark energy by
combining CMB, 2dFGRS and type-Ia supernova data. 
On the positive side, we find that by allowing the dark energy equation of state to
vary as a function of time, we do not introduce any strong
degeneracies that would adversely affect the standard cosmological parameter
estimation, apart from perhaps a mild degeneracy between the reionisation 
optical depth, $\tau$ and the redshift of the commencement of acceleration.
Its effect is to alter the ISW contribution to large angles and hence, after 
COBE normalisation uniformly alters the heights of the peaks \cite{BRUCE}. 
Breaking this degeneracy by some means will significantly enhance our ability 
to constrain dark energy dynamics. This might be done through a number of
routes: other astrophysical constraints on 
$\tau$; by probing the redshift of acceleration using ISW-LSS correlations;
by measuring the non-Gaussianity of CMB from weak lensing induced by structure
formation \cite{Fabio} or by beating 
cosmic variance using cluster polarisation at high redshift \cite{variance}.

The remaining cosmic parameters are only mildly affected by the new freedom
given to dark energy and are similar to their $\Lambda$CDM counterparts.
On the down side, with regard to the dark energy parameters, 
the currently available data set means that only the present value of the
equation of state $\w0$ 
is well constrained. We find $\w0<-0.8$ at 95\%
confidence level. By studying the behaviour of $w(z)$, we
conclude that the constraints become weak for any
redshifts larger than about unity. 

However, there are some clear results that emerge.
For example in the large class of Quintessence models that have 
periods of perfect tracking behaviour, i.e. $w=0$
during the matter dominated era, only those which
track for during the period $z\gsim 5$ are acceptable. 
Quintessence
models with a very late departure from tracking, or any dark
energy models with a late transition from a high value of $w$, are
disfavoured as a consequence of the fact that early contributions of 
dark energy are constrained to be negligible.

We have also found a point of practical importance in the hunt for dark energy
dynamics. Since only rapidly varying models at low redshift have a distinct 
signature as opposed to $\Lambda$CDM \cite{CORAS3} including such models in 
ones likelihood analysis is very important, otherwise the results will be biased towards 
no detection of dynamics. However these rapidly varying models are the most susceptible 
to numerical errors, which we have found can be as large as $5\%$. We therefore have used
a code specially adapted to handle these "kink" cases, and this is described in 
Appendix \ref{boltz}.  

We have also studied toy models for which $w<-1$ is possible.
In this case the perturbations are generically unstable, and
so we turned them off for any model which enters $w<-1$ at any
point. In this case models that have $w$ slightly larger than
$-1$ at early times and show a rapid, late transition to a
super-negative equation of state, $w\lsim-2$, are slightly,
but not significantly, preferred. In a slightly more physical
model where $w=-1$ cannot be crossed, allowing us to compute
the perturbations, we find no preference for these ``phantom'' models.

Overall, we conclude that dynamical models of dark energy
are by no means ruled out and provide a slightly better fit to
current data than \LCDM. The latter is still perfectly
acceptable and, given its simplicity, seems in many ways the
preferred model at the current time. Whether
observations and theoretical prejudice maintain this conclusio
over the coming years will have a profound impact on our understanding of
fundamental physics. 

\begin{acknowledgments}

We are grateful to Michael Doran for testing our models with CMBEASY and for the very useful discussions and suggestions.
P.S.C. thank Axel De La Macorra for the interesting discussions and the hospitality at the Mexico University.
P.S.C. is supported by Columbia University Academic Quality Fund,
M.K. and D.P. are grateful to PPARC for financial support and
B.A.B. is grateful to Z. Chacko 
for discussions and funding from Royal Society/JSPS.
Simulations were performed on COSMOS IV, the Origin 3800 supercomputer, funded by
SGI, HEFCE and PPARC. 
The completion of this paper was made possible thanks to the early
elimination of Italy, Switzerland and England from the European Championships.
We are grateful to the players of those teams for being so aware of our need
to finish this work.

\end{acknowledgments}

\appendix

\section{Parametrisation of the models\label{param}}
A simple way of describing a dark energy component is to consider a constant 
equation of state $w$.
However such an approach suffers from a number of drawbacks. In fact
assuming $w$ to be constant introduces a bias in the analysis 
of cosmological distance measurements
with the consequence that large negative values of $w$ are favoured 
if the dark energy is time dependent \cite{MAOR,VIREY}.
The effect of this bias has to be carefully taken into
account particularly when models of phantom energy \cite{CAL,CAR}, for which
$w<-1$, are constrained.
A possible way to address this issue is to constrain
a time dependent parametrisation of $w(z)$. 
In fact the time dependence of the dark energy equation of state completely specifies
the evolution of the dark energy density 
through the continuity equation.
Several formula have been proposed in the literature \cite{GERKE,WELLA,LINDER,PADMA}
all with limited applicability. They are typically based on Taylor expansions in some variable (e.g. $z$,
$\log (1+z)$ or $1-a$).

However, tracking quintessence models exhibit the important property
that before the universe begins to accelerate the dark energy has an
equation of state which mimics that of the dominant energy
component. Today of course
$w<-1/3$, so there must be a transition between the value at high redshifts and that
at the present time at some critical
redshift, $z_t$, or equivalently scale factor, $a_t = 1/(1+z_t)$, with
a thickness determined by a parameter $\Delta$. In fact, this
physically motivated parametrisation bears a striking resemblance to
the $\tanh(z)$ kink soliton solution in spirit. 

To compare various proposals we write it as:
\begin{equation}
w(a)=w_0+(w_m-w_0) \Gamma(a, a_t, \Delta);\label{bcc2}
\end{equation}
where $\Gamma$ determines precisely how the transition from $w=w_m$ to $w=w_0$ occurs. 
Two proposals for $\Gamma$ have been made in the literature. 
The original one\cite{BRUCE,CONDENSATION} uses $z$ instead of $a$:
\begin{equation}
\Gamma(z,z_t,\Delta) = \frac{-1}{1 + e^{\frac{z-z_t}{\Delta}}} 
\label{1trans}
\end{equation}
while a different proposal (and the one that we use here) was made in \cite{CORAS2}, namely: 
\begin{equation}
\Gamma(a,a_t,\Delta) = \frac{1+e^{\frac{a_t}{\Delta}}}{1+e^{-\frac{a-a_t}{\Delta}}}
\times\frac{1-e^{-\frac{a-1}{\Delta}}}{1-e^{\frac{1}{\Delta}}}
\label{2trans}
\end{equation}

The latter form has the advantage of being more stable numerically
(since $a$ is bounded in the interval $[0,1]$ while redshifts up to
$z=10^4$ must be considered in the first form) and satisfies
$w(a=1)=w_0$ whereas the same is only true for sufficiently rapid
transitions in eq. (\ref{1trans}). Because of this we use the latter
form (\ref{2trans}) in this paper. 

Importantly, this form can also be extended \cite{CORAS2} to allow for
a different value of $w$ during radiation domination so that it can
represent tracking models faithfully at all redshifts in
terms of five physical parameters: the value of the equation of state
today $w_0$, 
during the matter era $w_m$ and during the radiation era $w_r$,
while the time dependence is specified by the value of the scale factor
$a_t$ where the equation of state changes from $w_m$ to $w_0$ and
the width of the transition $\Delta$. Since Big-Bang Nucleosynthesis bounds
limit the amount of dark energy to be negligible during the radiation 
dominated era \cite{BEANMEL} the extra freedom is not particularly important and we further 
reduce our parameter space by setting $w_r=w_m$ yielding the transition function (\ref{2trans}). 

Written out fully then, the equation of state we use in this paper is Eq. (4) of ref. \cite{CORAS2}:
\begin{equation}
w(a)=w_0+(w_m-w_0)\times\frac{1+e^{\frac{a_t}{\Delta}}}{1+e^{-\frac{a-a_t}{\Delta}}}
\times\frac{1-e^{-\frac{a-1}{\Delta}}}{1-e^{\frac{1}{\Delta}}}.
\label{woa}
\end{equation}
The dark energy parameters specified by the vector $\overline{W}_{DE}=(w_0,w_m,a_t,\Delta)$
with the parametrisation given by Eq. (\ref{woa}) can account for most of the dark energy
models proposed in the literature (see \cite{CORAS2} for a detailed discussion). For instance,
models characterized by a slowly varying equation of state, such as supergravity inspired models
\cite{BRAX}, correspond to a region of our parameter space for which $0<a_t/\Delta<1$, while
models with a rapid variation of $w(a)$, such as the two exponential potential \cite{NELSON}
or the Albrecht-Skordis model \cite{ALBRECHT}, correspond to $a_t/\Delta>1$.
Models with a simple constant equation of state are given by $w_0=w_m$.
We can account also for the so called phantom energy 
models for which $w_0,w_m<-1$. The cosmological constant model corresponds
to the following cases: $w_0=w_m=-1$ or $w_0=-1$ and
$a_t\lesssim0.1$ with $a_t/\Delta>1$.

\section{Cosmological evolution of scalar fields\label{scaleq}}

The cosmological evolution of minimally coupled quintessence field Q
is described by the Klein-Gordon equation,
\begin{equation}
Q''+2\frac{a'}{a}Q'+a^2\frac{dV}{dQ}=0,\label{kg}
\end{equation}
the prime denotes derivatives with the respect to conformal time,  $V(Q)$ being the scalar field potential and
\begin{equation}
\left(\frac{a'}{a}\right)^2=\frac{8\pi G}{3}\left[\rho_m+\rho_r+\frac{{Q'}^2}{2a^2}+V(Q) \right],
\end{equation}
where $\rho_m$ and $\rho_r$ are the matter and radiation energy density
respectively. The equation of motion for the quintessence fluctuations at the scale $k$ in the synchronous gauge
is given by
\begin{equation}
\delta Q''+2\frac{a'}{a}\delta Q'+(a^2 \frac{d^2V}{dQ^2}+k^2)\delta Q=-\frac{1}{2}Q'h', \label{pertkg}
\end{equation}
where $h$ is the metric perturbation.
Instead of specifying the scalar field potential $V(Q)$ the conservation of the
energy momentum tensor allows us to describe a scalar field as a perfect
fluid with a time dependent equation of state  $w(a)$. 
In such a case the dark energy density evolves according to
\begin{equation}
\rho_{DE}(a)=\frac{3H_0^2}{8\pi G}\Omega_{DE} \exp{\left[-3 \int_a^1 \frac{1+w(s)}{s}ds\right]},
\end{equation}
where $H_0$ is the value of the Hubble parameter and $\Omega_{DE}$ is the dark energy density. 
We can describe the
perturbations in a dark energy fluid specified by $w(a)$ using Eq.~(\ref{pertkg}) with
the second derivative of the scalar field potential is 
written in terms of the time derivatives of the equation of state $w(a)$ \cite{BOB},
\begin{eqnarray}
a^2 \frac{d^2V}{dQ^2}=-\frac{3}{2}(1-w)\left[\frac{a''}{a}-\left(\frac{a'}{a}\right)^2\left(\frac{7}{2}+\frac{3}{2}w\right)\right] \nonumber \\
+\frac{1}{1+w}\left[\frac{{w'}^2}{4(1+w)}-\frac{w''}{2}+w'\frac{a'}{a}(3w+2)\right].\nonumber\\
\label{d2v2}
\end{eqnarray}
For models with an equation of state rapidly evolving towards $w=-1$ the Eq.~(\ref{d2v2}) is undefined
since $w=-1$ and $w',w''=0$ after the transition to $-1$. 
This corresponds to the fact that at
the classical level the vacuum state has no perturbations 
and quantities such as the sound speed are not
defined anymore. When such conditions are realized by one of our models in the MCMC chain,
we set the dark energy density perturbations to zero by the time at which $w=-1$.
On the other hand Eq.~(\ref{d2v2}) becomes singular at $w=-1$ if $w',w''\neq0$,
which could be the case for phantom dark energy models with $w(a)$ crossing the value $-1$. 
On the contrary perturbation in a time dependent phantom fluid with $w<-1$
at all the time can be described in our framework by switching the negative sign in
front of the second derivative of the potential in Eq.~(\ref{pertkg}).

\section{THE BOLTZMANN CODE - KINKFAST 1.0.0}\label{boltz}
We have modified the CMBFAST 4.5.1 to implement our parameterisation
of dark energy (KINKFAST 1.0.0). We have tested the numerical accuracy
of our code by comparing the power spectra of different dark energy
models with those computed by using CMBEASY \cite{CMBEASY}. 
\begin{figure}[t]
\begin{center}
\includegraphics[width=80mm]{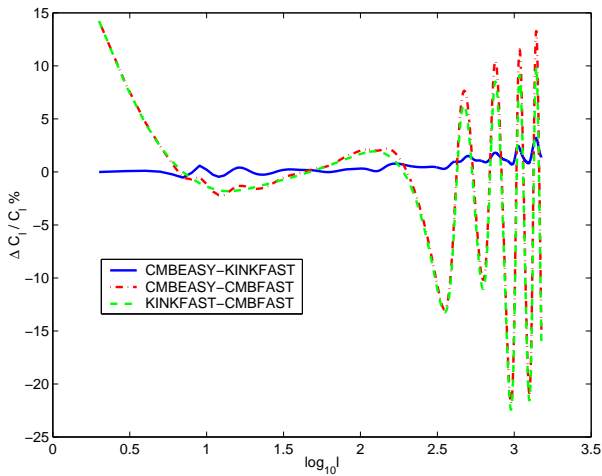} \\
\caption[cpar]{\label{rel} Relative difference of CMB spectrum for a rapidly varying dark energy model computed with
CMBEASY, KINKFAST 1.0.0 and CMBFAST 4.5.1.}
\end{center}
\end{figure}
We find a
perfect agreement within $1 \%$. On the other hand we notice that the
current version of CMBFAST 4.5.1, which implements the dark energy by
reading a redshift sampled equation of state, leads to wrong spectra
in the case of dark energy models with a rapid evolution in the
equation of state (see figure \ref{rel}). 
The cause of such a
discrepancy is that the dark energy density is obtained through a {\it
splint} integration procedure of the sampled equation of state. We
find this method to be of poor accuracy. Without an analytic
formula for $w(z)$, it is more reliable to derive a polynomial fitting
function for the sampled equation of state and integrate the
corresponding polynomial form with standard Numerical Recipes
\cite{num} integration subroutines such as {\it rombint}.

\end{document}